\documentclass[aps,preprint,prd,nofootinbib]{revtex4-1}
\usepackage{graphicx}
\usepackage{psfrag}

\usepackage{subfigure}
\usepackage{array}
\usepackage{graphicx}
\usepackage{amsmath}
\usepackage{amssymb}
\usepackage{mathrsfs}
\usepackage{bm}
\usepackage{slashed}
\usepackage{threeparttable}
\usepackage{multirow}
\usepackage[colorlinks, linkcolor=black, anchorcolor=black, citecolor=black]{hyperref}
\usepackage[amsmath,thmmarks,hyperref]{ntheorem}
\usepackage{appendix}
\usepackage{soul}

\allowdisplaybreaks[4]

\begin{document}

\title{Probing axion-like particles in leptonic decays of heavy mesons}
\author{Gang Yang$^1$\footnote{22s011005@stu.hit.edu.cn}, Tianhong Wang$^1$\footnote{thwang@hit.edu.cn (Corresponding author)}, and Guo-Li Wang$^{2}$\footnote{wgl@hbu.edu.cn}\\}
\address{$^1$School of Physics, Harbin Institute of Technology, Harbin, 150001, China\\
$^2$Department of Physics, Hebei University, Baoding 071002, China}

\baselineskip=20pt

\begin{abstract}

We study the possibility of finding the axion-like particles (ALPs) through the leptonic decays of heavy mesons. The Standard Model (SM) predictions of the branching ratios of the leptonic decays of heavy mesons are less than the corresponding experimental upper limits. This provides some room for the existence of decay channels, of which the ALP is one of the products.  Three scenarios are considered: First, the ALP is only coupled to one single charged fermion, namely, the quark, the antiquark, or the charged lepton; second, the ALP is only coupled to quark and antiquark with the same strength; and third, the ALP is coupled to all the charged fermions with the same strength. The constraints of the coupling strength in different scenarios are obtained by comparing the experimental data of the branching ratios of leptonic decays of $B^-$, $D^+$, and $D_s^+$ mesons with the theoretical predictions achieved by using the Bethe-Salpeter (BS) method. These constraints are further applied to predict the upper limits of the leptonic decay processes of the $B_c^-$ meson in which the ALP participates.

\end{abstract}

\maketitle

\section{Introduction}
Peccei and Quinn introduced a new global chiral symmetry~\cite{pec1977} in 1977, known as $U(1)_{\rm PQ}$ symmetry, to address the strong CP problem in QCD, that is, the absence of CP violation in the strong interactions and the electric dipole moment (EDM) of the neutron being forbidden. At energy scale $f_a$, such a symmetry is assumed to break spontaneously, resulting in the appearance of a pseudo-Nambu-Goldstone boson, namely the axion 
~\cite{pec1977,pq1977,Wil1977,wei1978,dil2020}, whose mass $m_a$ is constrained by the relation $m_a \sim \frac{m_\pi f_\pi}{f_a}$, where $m_\pi$ and $f_\pi$ are the mass and decay constants of pion, respectively. This means that if the energy scale $f_a$ is very high, the axion should be extremely light. The condition can be relaxed if we are not limited to such a QCD axion but consider more general axion-like particles (ALPs)~\cite{baum2017,bri2017,mar2020}. In such cases, both the PQ symmetry breaking scale $f_a$ and the ALP mass $m_a$ can be considered as independent parameters.

People are interested in ALPs in many aspects. Theoretically, several ALPs are predicted by string theories~\cite{Arv2009,Cic2012} and supersymmerty\cite{nel1994,bag1994}. These particles may be crucial to the evolution of cosmology, most importantly, as a candidate of dark matter~\cite{Marsh2015}. The cosmological and astronomical observations can set strict constraints on very light ALPs  \cite{Ira2011,aya2014,vin2016,gao2023,bud2014,ake2017,aga2022}.  The phenomenology of ALPs have also been extensively studied at the Large Hadron Collider (LHC)~\cite{Alo2023,Bie2022,bau2017,jae2015,ATLAS2023,Buo2023,Ghe2022,Fen2018,Bis2023}. For example, they can be produced by the decays of on-shell Higgs/$Z$ boson~\cite{Bis2023,baum2017}, or participate as off shell mediators in the scattering processes~\cite{Gavela2019}. At high-luminosity electron-positron colliders, ALPs with masses in the MeV-GeV range can also be explored non-resonantly or resonantly~\cite{aca2024,zha2023,fer2023,bel2020,mer2019,Bon12022,Bere1989}, which means they can be produced directly by coupling to the charged leptons or the gauge bosons, or produced through the decays of final mesons. For example, in Ref.~\cite{mer2019} the production of ALPs at $B$ factories is considered, including the contributions of $e^{+}e^{-} \to \gamma a$ and  $e^+e^-\to\Upsilon\to\gamma a$. 
 Recently, the BESIII experiment~\cite{besiii2023} use a data sample of $\psi(3686)$ to obtain the upper limits of the branching fraction of the decay $J/\psi \to \gamma a$ and set constrains on the coupling $g_{a\gamma\gamma}$ in the mass range of $0.165~{\rm GeV} \leq m_a \leq 2.84~{\rm GeV}$. The fixed target experiments are also promising methods for searching for ALPs~\cite{ber2024,ema2023,afi2023,NA2023,Col2022,Dob2018,Har2019}, as the detectors can extend tens of meters and are suitable for detecting long-lived particles. ALP production through $K$ meson decays are usually investigated in such cases ~\cite{ber2024,ema2023,afi2023,NA2023}.  

\begin{figure}[htbp]
	\centering
	\graphicspath{{figures/}}
	\includegraphics[width=2.5in,height=1in]{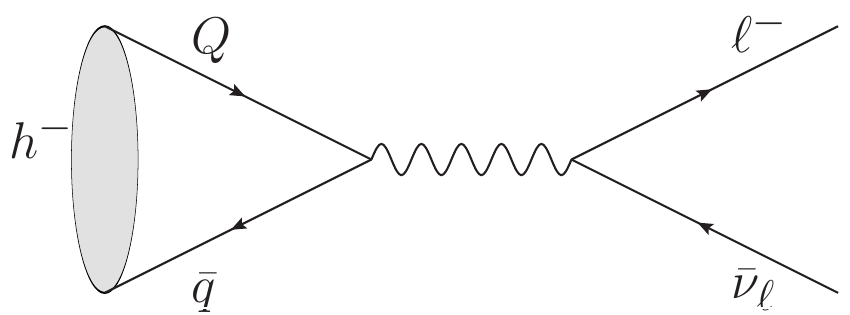}
	\caption{Diagram of two-body leptonic decays of a charged heavy meson.}
	\label{t0}
\end{figure}

Except for the methods mentioned above, the decays of heavy-flavored mesons also provide a way to probe ALPs. On one hand, a large number of bottomed and charmed mesons have been produced at $B$ factories and the $\tau$-charm factory, respectively, and on the other, a larger range of $m_a$ can be looked into in heavy meson decays compared with that of $K$ meson. We will focus on the decay processes $h^- \to \ell^- \slashed E$, where $h^-=D^-,~D_s^-, ~B^-$, or $B_c^-$; $\ell^-=e^-,~\mu^-$, or $\tau^-$; $\slashed E$ represents the missing energy. In the Standard Model, $\slashed E$ is transported by the antineutrino (see Fig.~\ref{t0}), and the corresponding partial width is
\begin{equation}\label{1.3}
	\Gamma\left(h^-\rightarrow \ell^-\bar{\nu}_{\ell}\right)=\frac{G_F^2}{8\pi}\left|V_{Qq}\right|^2f_h^2M^3\frac{m_\ell^2}{M^2}\left(1-\frac{m_\ell^2}{M^2}\right)^2,
\end{equation}
where $G_F$ is the Fermi coupling constant, $f_h$ is the decay constant of $h$ meson, $V_{Qq}$ is the CKM matrix element, $M$ and $m_\ell$ are the masses of the meson and the lepton, respectively, and the neutrino mass is assumed to be zero. If an ALP can also be produced and has sufficient lifetime to escape the detector, we will have $\Gamma(h^-\to \ell^-\slashed E)=\Gamma(h^-\to \ell^-\bar{\nu}_{\ell})+\Gamma(h^-\to \ell^-\bar{\nu}_{\ell}a)$. Then the experimental results of $h^-\to \ell^- \slashed E$, which are presented in Table 1, can be used to constrain the coupling between the ALP and SM particles.
\begin{table}[htbp]\centering
	\setlength{\tabcolsep}{12pt}
	\caption{Experiment results of ${\mathcal B}(B^- \to \ell^-\slashed E)$, ${\mathcal B}(D^-\to \ell^-\slashed E)$, and ${\mathcal B}(D_s^-\to \ell^-\slashed E)$.}\label{data}
	\scalebox{1.0}{
		\begin{tabular}{cc}
			\hline
			Channel & Experiment values\\
			\hline
			${\mathcal B}(B^-\to e^- \slashed E)$ & $\le 9.8 \times 10^{-7}$~\cite{Belle2006}\\
			${\mathcal B}(B^-\to \mu^- \slashed E)$& $(5.3 \pm 2.0 \pm 0.9) \times 10^{-7}$ ~\cite{Belle2019}\\
			${\mathcal B}(B^-\to \tau^- \slashed E)$&$(7.2 \pm 2.7 \pm 1.1 ) \times 10^{-5}$~\cite{Belle2012}\\
			${\mathcal B}(D_{s}^-\to e^- \slashed E)$  & $ \le 0.83 \times 10^{-4}$  ~\cite{Belle2013}\\
			${\mathcal B}(D_{s}^-\to \mu^- \slashed E)$& $(0.5294 \pm 0.0108 \pm 0.0085) \times 10^{-2}$~\cite{2023cym}\\
			${\mathcal B}(D_{s}^-\to \tau^- \slashed E)$&$(5.44 \pm 0.17 \pm 0.13 ) \times 10^{-2}$~\cite{2023fhe}\\
			${\mathcal B}(D^-\to e^- \slashed E)$ & $ \le 8.8 \times 10^{-6}$~\cite{CLEO2008} \\
			${\mathcal B}(D^-\to \mu^- \slashed E)$& $(3.71 \pm 2.7 \pm 1.1) \times 10^{-3}$ ~\cite{abl2014}\\
			${\mathcal B}(D^-\to \tau^- \slashed E)$&$(1.20 \pm 0.24 \pm 0.12 ) \times 10^{-5}$~\cite{besiii2019}\\
			\hline
		\end{tabular}
	}
\end{table}

This study is inspired by Refs.~\cite{adi2012,gue2022,gue2023,gal2022} where the processes $h^- \to \ell^- \bar{\nu}_{\ell}a$ with $a$ being an ALP were studied. Beyond the considerations for the case of a pseudoscalar, Ref.~\cite{adi2012} also considered the possibility of $a$ being a vector particle. Here, we will also probe such processes, but with three main differences. First, the Hadron transition matrix elements will be calculated using the Bethe-Salpeter (BS) method. This method has been widely used to study the weak decays~\cite{Zhou2020,fu2011,zha2010}, strong decays~\cite{2021dwh}, and electromagnetic decays~\cite{Wan2019} of heavy mesons,	
and the theoretical results are consistent with the experimental data. Second, for the coupling of ALP with quarks, we consider the interference effects of different diagrams and extract the allowed range of parameters in a more general way. Third, we will explore ALP production from the decays of the double heavy meson $B_c$, that is $B_c^- \to \ell^- \bar{\nu}_{\ell}a$, which could give the constraint in a larger ALP mass range.

The remainder of the paper is organized as follows: In Sec.~II, we outline the effective Lagrangian, which describes the interaction between ALPs and charged fermions. Then, we calculate the Hadronic transition matrix elements of the $h^-\to \ell^-\nu_{\ell} a$ processes using the BS method. In Sec.~III, we apply the experimental results to extract the allowed regions of the parameters, which are then used to calculate the upper limits of the branching fractions of the $B_c$ decay processes. Finally, we present the summary in Sec.~IV.


\section{Theoretical Formalism}
\label{sec2}

\begin{figure}[!h]
	\centering
	\graphicspath{{figures/}}
	\subfigure[]{
		\includegraphics[width=2.5in,height=1in]{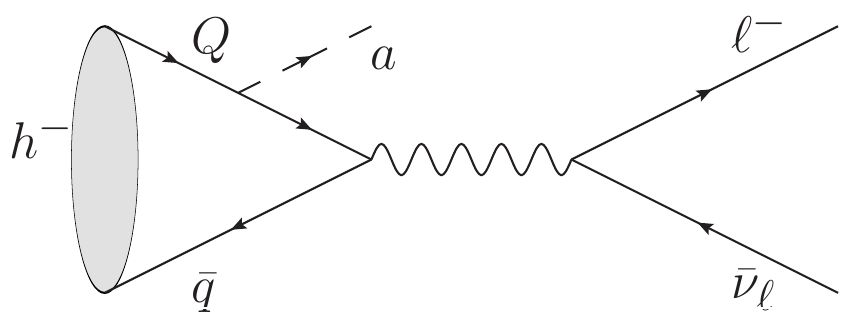}\label{1a}
	}
	\quad
	\subfigure[]{
		\includegraphics[width=2.5in,height=1in]{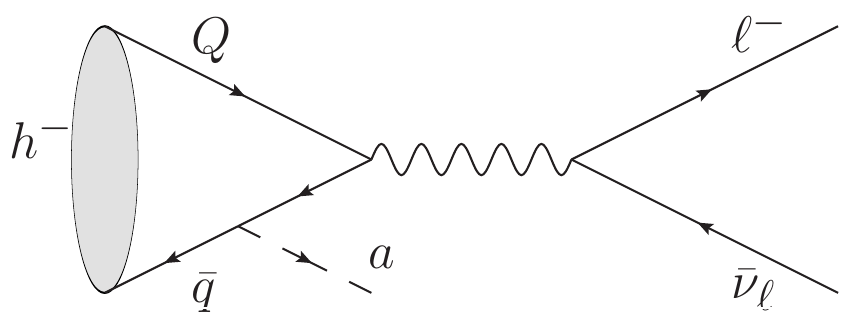}\label{1b}
	}
	\quad
	\subfigure[]{
		\includegraphics[width=2.5in,height=1in]{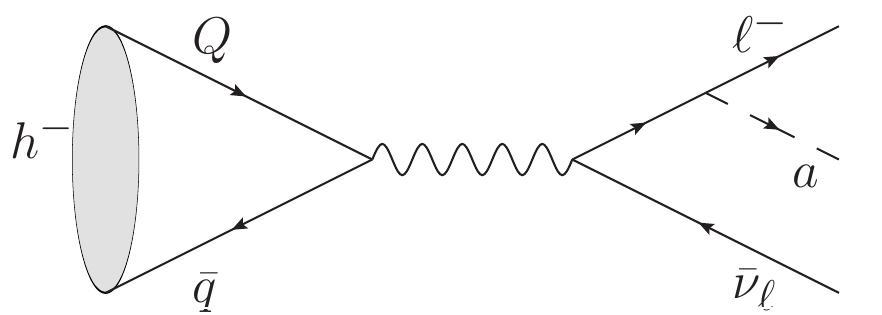}\label{1c}
	}
	\quad
	\subfigure[]{
		\includegraphics[width=2.5in,height=1in]{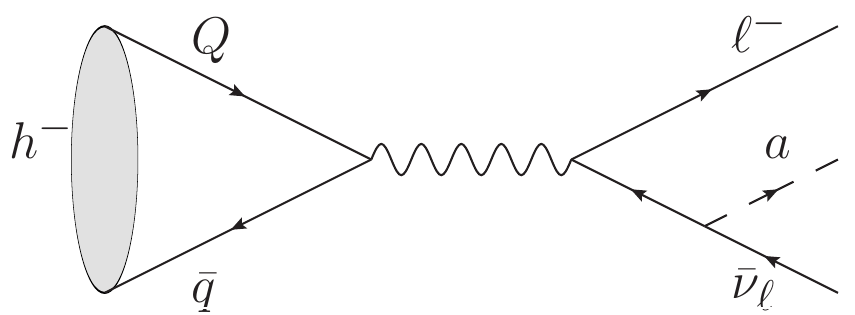}\label{1d}
	}
	\caption{Feynman diagrams of $h \rightarrow \ell\bar{\nu}_\ell a$. $(a)$, $(b)$, $(c)$, and  $(d)$ represent the ALP couples with the quark, antiquark, charged lepton, and antineutrino, respectively.}
	\label{fig1}
\end{figure}

Generally, the ALPs can interact with the fermions and gauge bosons. Herein, in the case of leptonic decays of heavy mesons, we only consider the tree level diagrams shown in Fig.~\ref{fig1}, which represent the interaction of the ALPs with SM fermions. The contributions of gauge bosons are of a higher order and assumed to be neglected. The effective Lagrangian, which describes flavor conserving processes, comprises dimension-five operators~\cite{adi2012,gue2022,gue2023,gal2022}
\begin{equation}\label{eq2.1}
	\mathcal{L}_{eff} \supset - \frac{\partial_{\mu} a}{2 f_{a}} \sum_{i } c_{i} \bar{\psi_{i}} \gamma^{\mu} \gamma_{5} \psi_{i} \Rightarrow  i \frac{a}{f_a} \sum_{i } c_i m_i \bar{\psi_{i}} \gamma_{5} \psi_{i},
\end{equation} 
where $c_i$ is the effective coupling constant with the index $i$ extending over all the fermions. Similar with Refs.~\cite{gue2022,gue2023,gal2022}, in the last step, we have used the equation of motion to rewrite the expression. This corresponds to the use of another operator basis, which will lead to different upper limits of $c_i$. However, when we calculate the decay channels of $B_c$ meson, two methods give quite similar results. The dependence of the fermion mass $m_i$ indicates that the ALP does not couple directly with the neutrinos and Fig.~2(d) does not need to be considered, which will simplify the calculation. 

The Feynman amplitude corresponding to each diagram can be written using the meson wave function. Here, the instantaneous BS wave function is appropriate for the calculation, as we are considering the heavy flavored meson, which can be seen as a two-body bound state including at least one heavy quark (or antiquark). This kind of wave function has been extensively used to study the decay processes of heavy mesons. For the pseudoscalar meson, its wave function has the form~\cite{kim2003}
\begin{equation}\label{eq2.3}	
	\varphi\left(q_\perp\right)=M \left[\frac{\slashed P}{M}f_{1}(q_{\perp})+f_{2}(q_\perp)+\frac{\slashed q_\perp}{M}f_{3}(q_{\perp})+\frac{\slashed P \slashed q _\perp}{M^2}f_{4}(q_\perp)\right]\gamma^5,
\end{equation}
where $P$ is the meson momentum and $q$ is the relative momentum between the quark and antiquark; $q^\mu_\perp \equiv q^\mu- \frac{P \cdot q }{M^2} P^\mu = (0,~\vec q)$; $f_i$ is the radial wave function, which can be achieved by solving the eigenvalue equations numerically~({\sl e.g.}, the detailed results for $B_c$ meson can be found in Ref.~\cite{2022wang}). 

The amplitude corresponding to Fig.~2(a) can be written as 
\begin{equation}\label{eq2.16}
	\begin{aligned}
		\mathcal{M}_{a}
		&=-i\frac{G_F}{\sqrt{2}} \frac{c_{Q}m_{Q}}{f_{\text{a}}}V_{qQ}\sqrt{N_c}\int\frac{d^{3} q_{\perp}}{{\left(2\pi\right)}^{3}}\text{Tr}\left[\gamma_{\mu}\left(1-\gamma^{5}\right)\frac{1}{\slashed p_Q-\slashed k_a-m_Q}\gamma^{5}\varphi\left(q_\perp\right)\right]\\
		&\times\bar{u}_{\ell}(k_\ell)\gamma^{\mu}\left(1-\gamma^{5}\right)v_{\bar\nu}(k_{\bar\nu}),
	\end{aligned}
\end{equation}
where $c_Q$ is the coupling constant; $m_Q$ is the mass of quark Q; $N_c=3$ is the color factor; $k_a$, $k_\ell$, and $k_{\bar\nu}$ are the momenta of ALP, charged lepton, and antineutrino, respectively; $p_Q$ is the momentum of quark Q, which is related to $P$ and $q$, as $p_Q = \frac{m_Q}{m_Q+m_q} P+q$, with $m_q$ being the mass of the antiquark. To obtain Eq.~(4), we have made an approximation, namely $p_Q \approx \frac{m_Q}{m_Q+m_q} P+q_\perp$, so that the denominator is independent of the time component of the relative momentum $q$.
After integrating out $q_\perp$, $\mathcal{M}_a$ becomes
\begin{equation}\label{eq2.17}
	\begin{split}
		\mathcal{M}_{a}
		= -i A_{Q}\left(F_{1}P^{\mu}+F_{2}k_a^{\mu}\right)\bar{u}_{\ell}(k_\ell)\gamma^{\mu}\left(1-\gamma^{5}\right)v_{\bar\nu}(k_{\bar\nu}),
	\end{split}
\end{equation}
where we have used $A_Q = \frac{c_Q m_Q G_F V_{qQ}}{f_a \sqrt{2}}$ for short; $F_1$ and $F_2$ are the form factors of the hadronic transition matrix element, which are functions of the squared momentum transition $(P-k_a)^2$.

Similarly, the Feynman amplitude corresponding to Fig.~2(b) can be written as
\begin{equation}\label{eq2.7}
	\begin{aligned}
		\mathcal{M}_{b}
		& = i\frac{G_F}{\sqrt{2}}\frac{c_{q}m_{q}}{f_{a}}V_{qQ}\sqrt{N_c}\int\frac{d^{3} q_{\perp}}{{\left(2\pi\right)}^{3}}\text{Tr}
		\left[\gamma_{5}\frac{1}{\slashed p_{q}-\slashed k_a + m_{q}}\gamma^{\mu}\left(1-\gamma^{5}\right)\varphi\left(q_\perp\right)\right]\\
		&\times\bar{u}_\ell(k_\ell)\gamma_{\mu}\left(1-\gamma^{5}\right)v_{\bar\nu}(k_{\bar\nu}),
	\end{aligned}
\end{equation}
where $c_q$ is the coupling constant; $m_q$ and $p_q$ are the mass and momentum of the antiquark, respectively; and $p_q$ has the form $p_q\approx \frac{m_q}{m_Q+m_q} P-q_\perp$. After integrating out $q_\perp$, we get
\begin{equation}\label{eq2.10}
	\mathcal{M}_b=iA_q (G_{1}P^{\mu}+G_{2}k^{\mu})\bar{u}_\ell(k_\ell)\gamma_{\mu}\left(1-\gamma^{5}\right)v_{\bar\nu}(k_{\bar\nu}),
\end{equation}
where $A_q=\frac{c_q m_q G_FV_{qQ}}{f_a \sqrt{2}}$; $G_1$ and $G_2$ are form factors.

When calculating the integral, one must be careful, since as the relative momentum $\vec q$ changes, the denominator of the propagator may be zero, which will lead to a nonvanishing imaginary part of the form factors~\cite{2015Ju}. Specifically, we can write the propagator as 
\begin{equation}\label{eq2.8}
	\begin{split}
		\frac{\text{1}}{\slashed p_{Q/q}-\slashed k_a+m_{Q/q}-i\epsilon} =\frac{\slashed p_{Q/q}-\slashed k_a-m_{Q/q}}
		{a+b\cos\theta +i\epsilon},
	\end{split}
\end{equation}
where $\theta$ is the angle between $\vec q$ and $\vec k_a$. $a$ and $b$ are expressed as
\begin{equation*}\label{eq2.9}
	\begin{split}
		& a=\left(\frac{m_{Q/q}}{m_Q+m_q}M-E_a\right)^2-m_{Q/q}^2-\vec q^2-\vec k_a^2,\\
		& b=\pm 2 |\vec{q}||\vec{k}|,
	\end{split}
\end{equation*}
where $E_a$ is the energy of the ALP. We will first integrate out $\cos\theta$ analytically, and then integrate out $|\vec q|$ numerically. As $\vec q$ changes, $a+b$ and $a-b$ may have opposite signs, which means the pole can exist with a specific value of $\theta$.

The Feynman amplitude corresponding to Fig.~2(c) can be written as
\begin{equation}\label{eq2.2}
	\begin{split}
		\mathcal{M}_{c} & =-i\frac{G_F}{\sqrt{2}}\frac{c_{\ell}m_{\ell}}{f_a}V_{qQ}\sqrt{N_c}\int\frac{d^{3} q_{\bot}}{{\left(2\pi\right)}^{3}}
		\text{Tr}\left[\gamma^{\mu}\left(1-\gamma^{5}\right) \varphi\left(q_\perp\right)\right]  \\
		& \times\bar u_\ell(k_\ell)\gamma_{5}\frac{1}{\slashed k_{\ell}+\slashed k_a-m_{\ell}}\gamma_{\mu}\left(1-\gamma^{5}\right)v_{\bar\nu}(k_{\bar\nu}).
	\end{split}
\end{equation}
where $c_\ell$ is the coupling constant. The hadronic part of the amplitude is proportional to the meson momentum~\cite{cve2004}, with the proportionality being the decay constant $f_h$
\begin{equation}\label{eq2.5}
   \sqrt{N_c}\int \frac{d^3q_\perp}{(2\pi)^3}\mathrm{Tr} \left[ \gamma^{\mu} \left(1-\gamma_{5}\right)\varphi(q_\perp)\right] = f_h P^{\mu}.
\end{equation}
Using Eq.~(10) and defining $A_\ell = \frac{c_\ell m_\ell G_FV_{qQ}}{f_a \sqrt{2}}$ for short, $\mathcal{M}_c$ is simplified to
\begin{equation}\label{eq2.6}
	\begin{split}
		\mathcal{M}_{c} =
		-i A_{\ell}f_h \bar{u}_\ell(k_\ell)\gamma_{5}\frac{\slashed p_{\ell}+\slashed k_a+m_{\ell}}{m^{2}_a+2 p_\ell\cdot k}\slashed P\left(1-\gamma^{5}\right)v_{\bar\nu}(k_{\bar\nu}).
	\end{split}
\end{equation}

The total Feynman amplitude is $\mathcal M=\mathcal M_a +\mathcal M_b +\mathcal M_c$. The partial width of such a decay channel is achieved by finishing the three-body phase space integral
\begin{equation}\label{eq2.24}
	\Gamma = \frac{1}{64 \pi^3 M} \int_{m_\ell}^{E_{\ell max}} dE_\ell \int_{E_{\bar\nu min}(E_\ell)}^{E_{\bar\nu max}(E_\ell)} dE_{\bar\nu} \overline{|\mathcal{M}|^2},
\end{equation}
where the upper and lower limits have the following forms:
\begin{equation*}
	\begin{aligned}
		 E_{\ell max}&=\sqrt{\left(\frac{M^2+m_\ell^2-m_a^2}{2M}\right)^2-m_\ell^2},\\
		 E_{\bar\nu max/min}(E_\ell)&= \frac{M^2-2ME_\ell+m_\ell^2-m_a^2}{2(M-E_\ell\mp\sqrt{E_\ell^2-m_\ell^2})}.
	\end{aligned}
\end{equation*}


\section{Numerical Results}

In this section, we will first use the formulas obtained above to calculate the limits of the coupling constants, and then the results will be applied to investigate the similar decays of the $B_c$ meson. We will focus on three different scenarios: First, the ALP couples only with a single fermion, namely the quark, the antiquark, or the charged lepton; second, the ALP couples with quarks and antiquarks but not with leptons; third, the ALP couples with all the charged fermions with the same coupling constant.

During the calculation process, the following numerical values of three groups of physical quantities are adopted: (1) The masses of constituent quarks are from Ref.~\cite{2017li}: $m_u = 0.305$ GeV, $m_d = 0.311$ GeV, $m_s = 0.5$ GeV, $m_c = 1.62$ GeV, and $m_b = 4.96$ GeV; (2) The relevant CKM matrix elements are from Particle Data Group (PDG)~\cite{wor2022}: $|V_{ub}| = 3.82 \times 10^{-3}$, $|V_{cd}| = 0.211$, $|V_{cs}| = 0.975 $, and $|V_{cb}| = 41.1 \times 10^{-3}$; (3) The decay constant of $D$, $D_s$, and $B$ mesons are the Lattice QCD results~\cite{FLAG2022}:  $f_D = 212.6$~MeV, $f_{D_s}=249.9$~MeV, and $f_B = 188$~MeV.  

\subsection{Scenario 1}

\begin{figure}[htbp]
	\centering
	\graphicspath{{figures/}}
	\subfigure[]{
		\includegraphics[width=2.5in,height=2.5in]{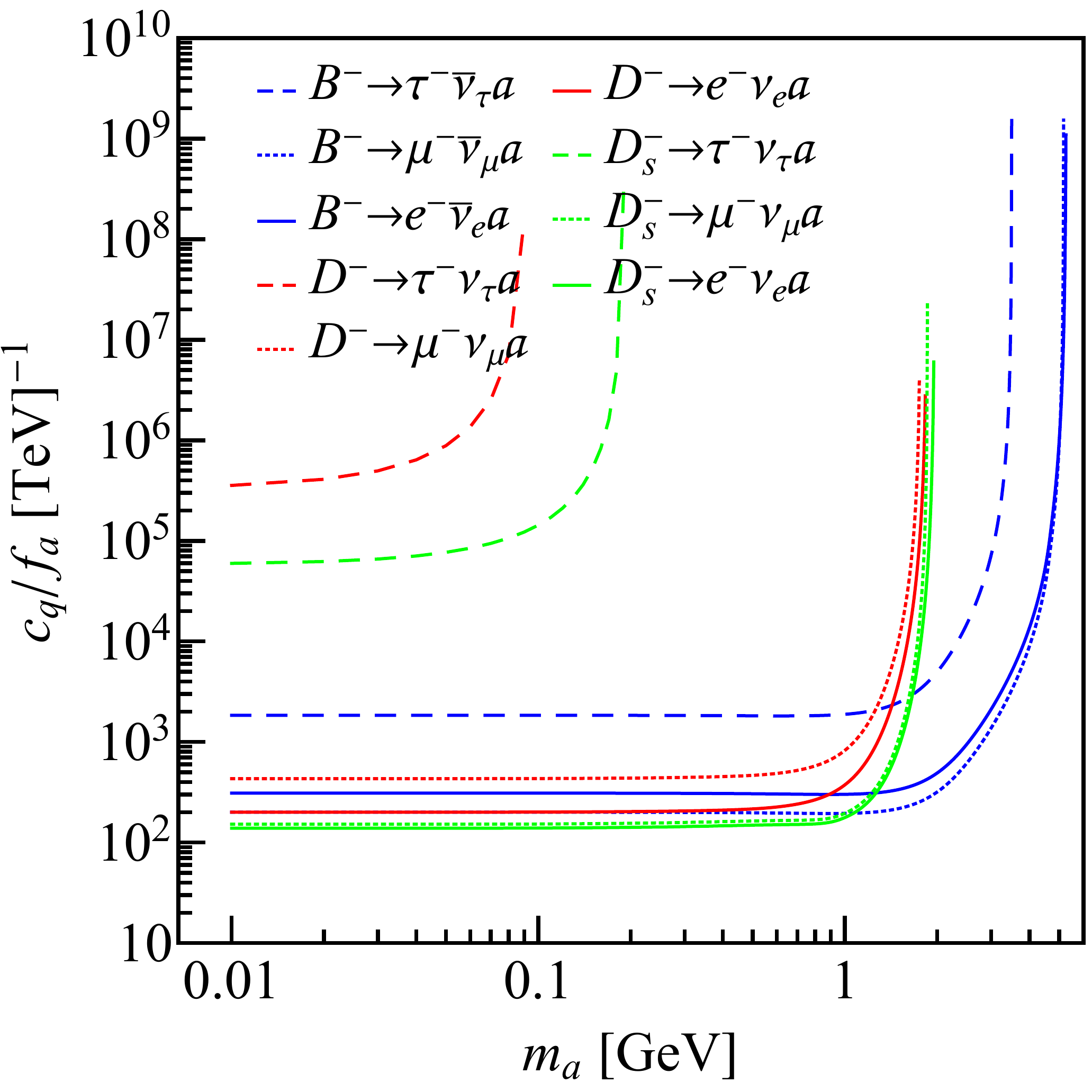}\label{a1}
	}\quad
	\subfigure[]{
		\includegraphics[width=2.5in,height=2.5in]{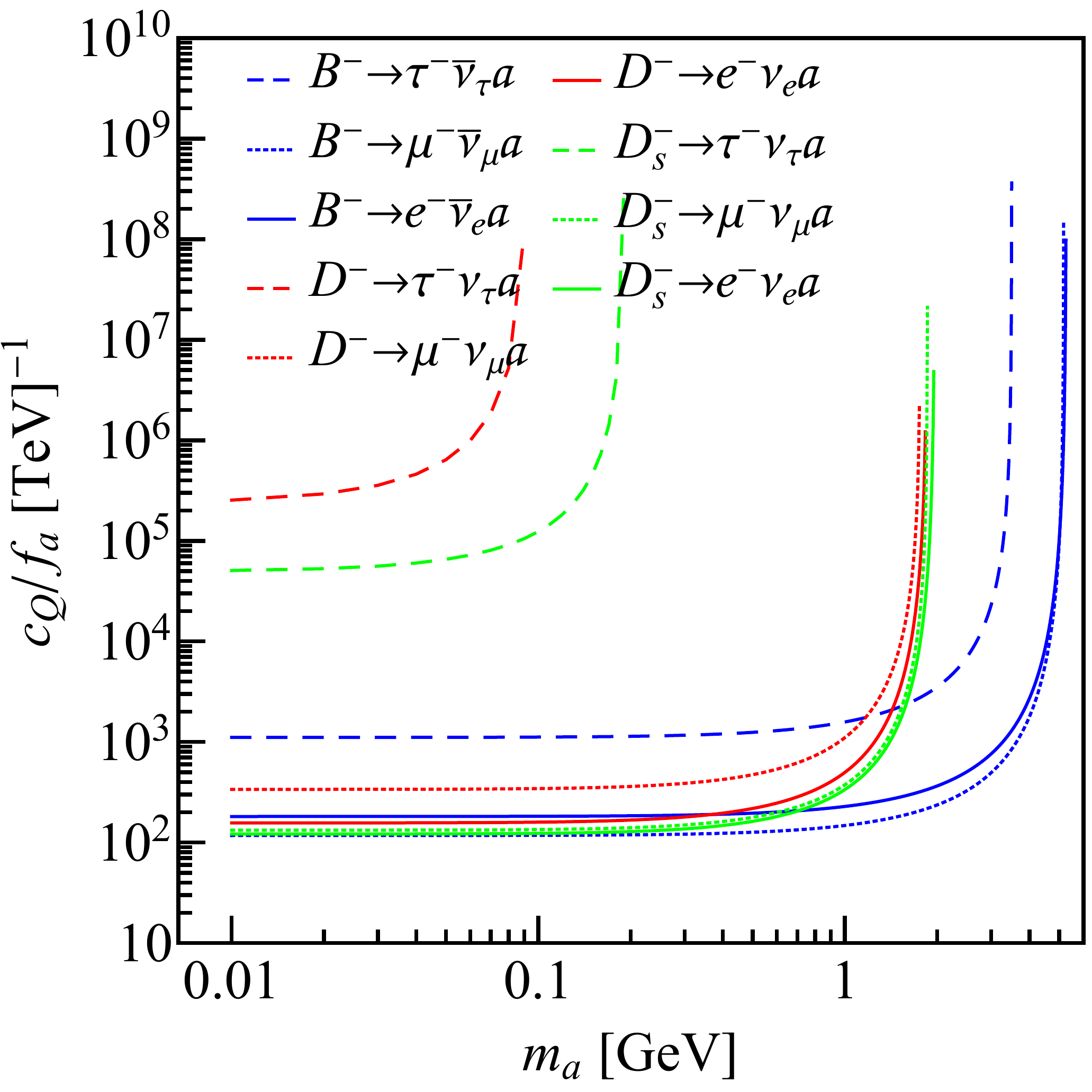}\label{b1}
	}\quad
	\subfigure[]{
		\includegraphics[width=2.5in,height=2.5in]{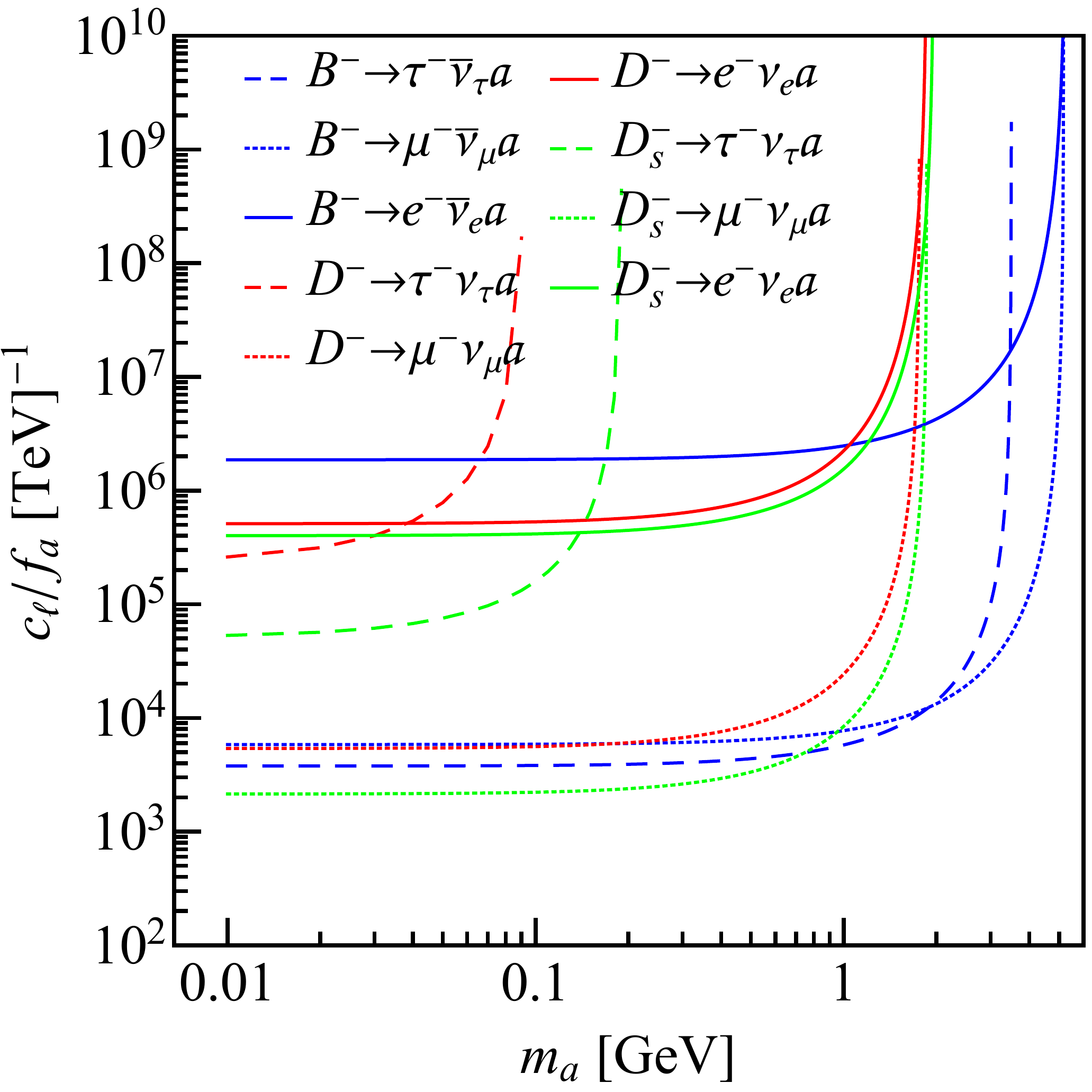}\label{c1}
	}\quad
	\caption{(color online) The upper limits of the coupling constants (a) $c_Q / f_a$ , (b) $c_q / f_a$ and (c) $c_{\ell} / f_a$ derived from the leptonic decays of $D$, $D_s$, and $B$ mesons. }
	\label{g1}
\end{figure}

We separately consider the coupling between the ALP and each charged fermion. For example, by setting $c_q$ and $c_\ell$ to zero, we get the branching fraction of $h^-\to\ell^-\bar\nu a$ with $c_Q/f_a$ and $m_a$ being unknown parameters. Then, using the experimental results in Table I, we get the upper limits of $c_Q/f_a$ as functions of $m_a$. The same is true for the other two cases. The results are shown in Fig.~\ref{g1}. One notices that the curves raise rapidly when $m_a$ is sufficiently, which is mainly owing to the reduced phase space. The most stringent upper limits for different coupling constants result from different decay channels. For example, Fig.~3(a) shows that the most strict constraints of $c_{b}/f_a$ and $c_{c}/f_a$, which are of the order of $10^2~{\rm TeV}^{-1}$ when $m_a$ is less than $1$~GeV, result from  $B^-\to\mu^-\bar\nu_\mu a$ and $D_s^+\to e^+\nu_e a$, respectively. One can also notice that the upper limit of $c_e/f_a$ in Fig.~3(c) is much higher than those of $c_{Q(q)}/f_a$ set up by the same channels, which is because of the small $m_e$. 

Using the above results, we can impose restrictions on $\mathcal B(B_c^-\to \ell^- \bar{\nu}_\ell a)$, which are presented in Fig. 4. We can see that when $m_a$ is less than 1~GeV, the branching ratio upper limits for ALP-quark coupling cases are of the order of $10^{-4}$. For the ALP-tau coupling case, the order of magnitudes is $10^{-2}$, which is much larger than those of the ALP-$e/\mu$ coupling cases. We also notice that all the curves, except the one corresponding to $B_c^-\to\tau^-\bar\nu_\tau a$ in Fig.~4(a), show an ascending trend. This is because the maximum value of $m_a$ permitted in the $B_c$ decay is larger than that in the same decays of the $B$ and $D_{(s)}$ mesons. For example, when we use $c_{b}/f_a$ from $B^-\to\mu^-\bar\nu_\mu a$ to constrain the branching ratios of $B_c^-\to\mu^-\bar\nu_\mu a$, the result blows up before $m_a$ reaches its maximum. The appearance of kinks in Fig.~4(c) is because we used $c_{e/\mu}/f_a$ from different channels when $m_a$ takes different values. 
\begin{figure}[htbp]
	\centering
	\graphicspath{{figures/}}
		\subfigure[~ALP couples with $b$ quark]{
		\includegraphics[width=2.5in,height=2.5in]{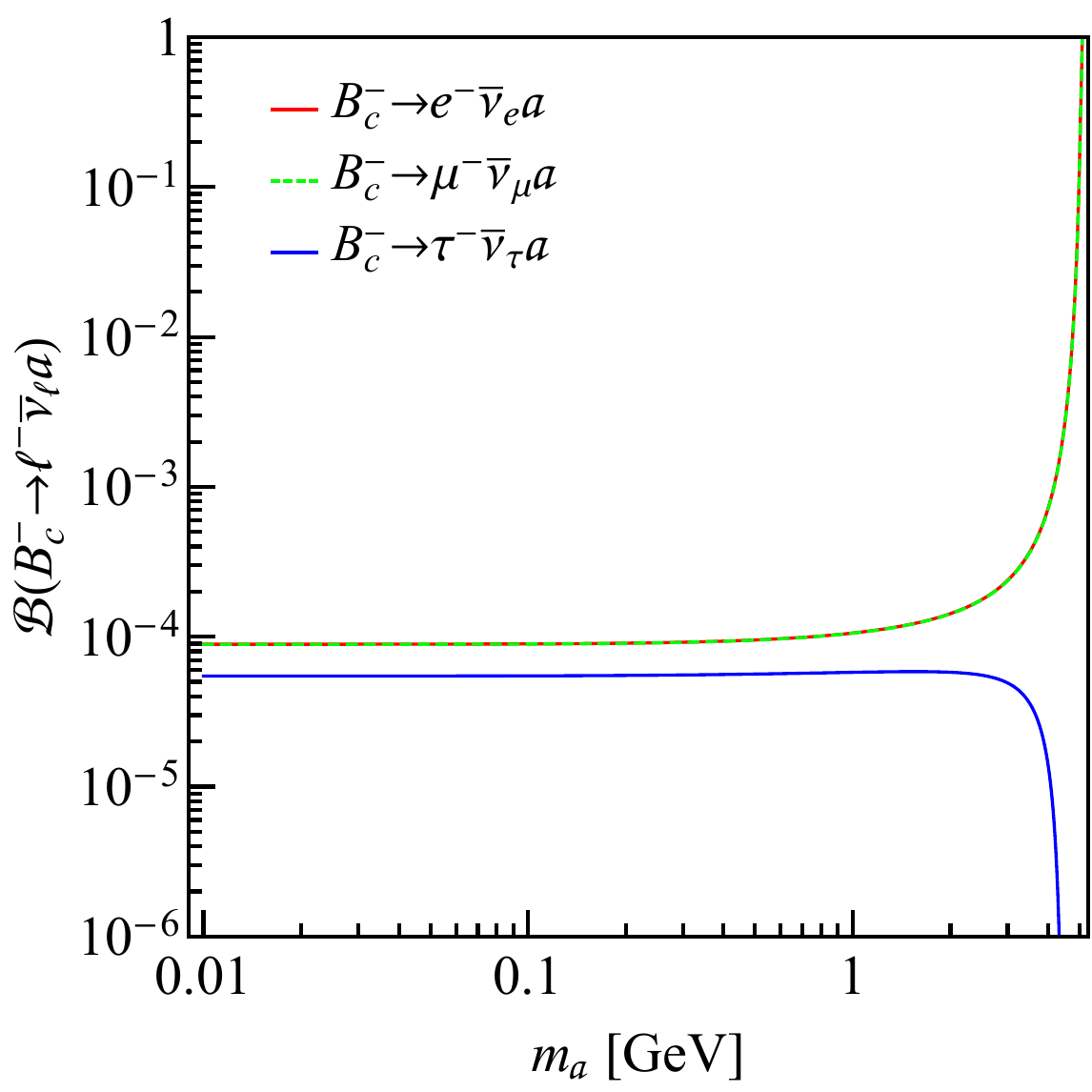}\label{a}
	}
	\quad
	\graphicspath{{figures/}}
	\subfigure[~ALP couples with $c$ quark]{
		\includegraphics[width=2.5in,height=2.5in]{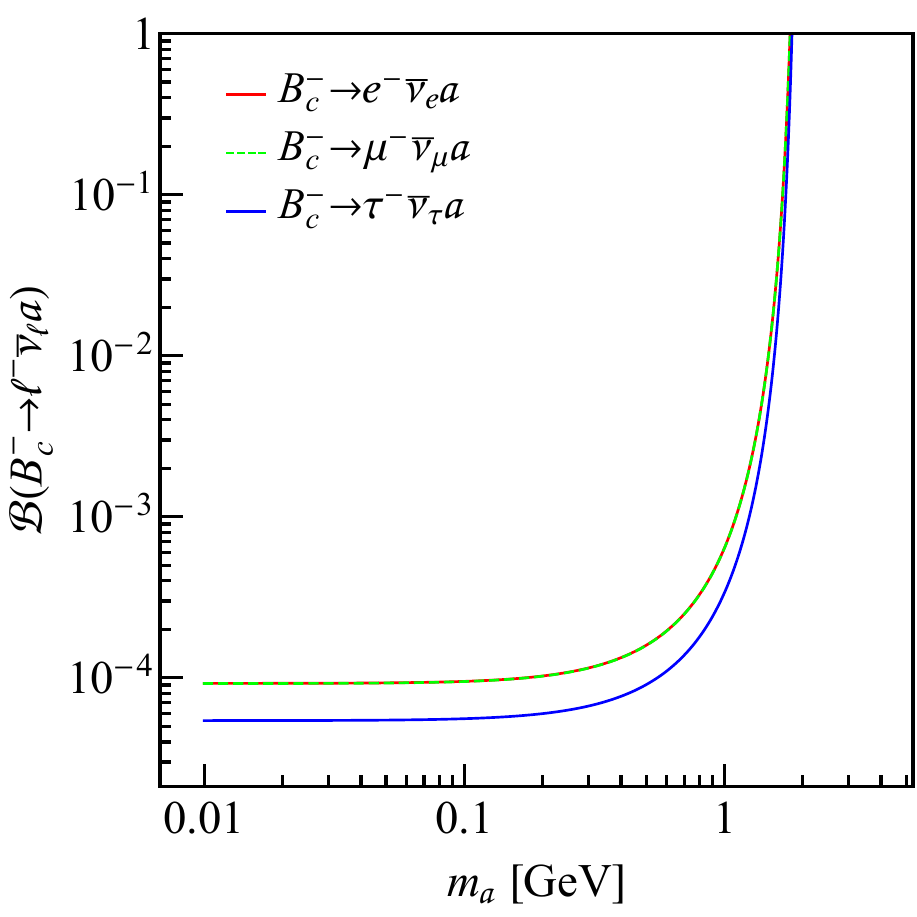}\label{b}
	}
	\quad
	\subfigure[~ALP couples with $\ell^-$]{
		\includegraphics[width=2.5in,height=2.5in]{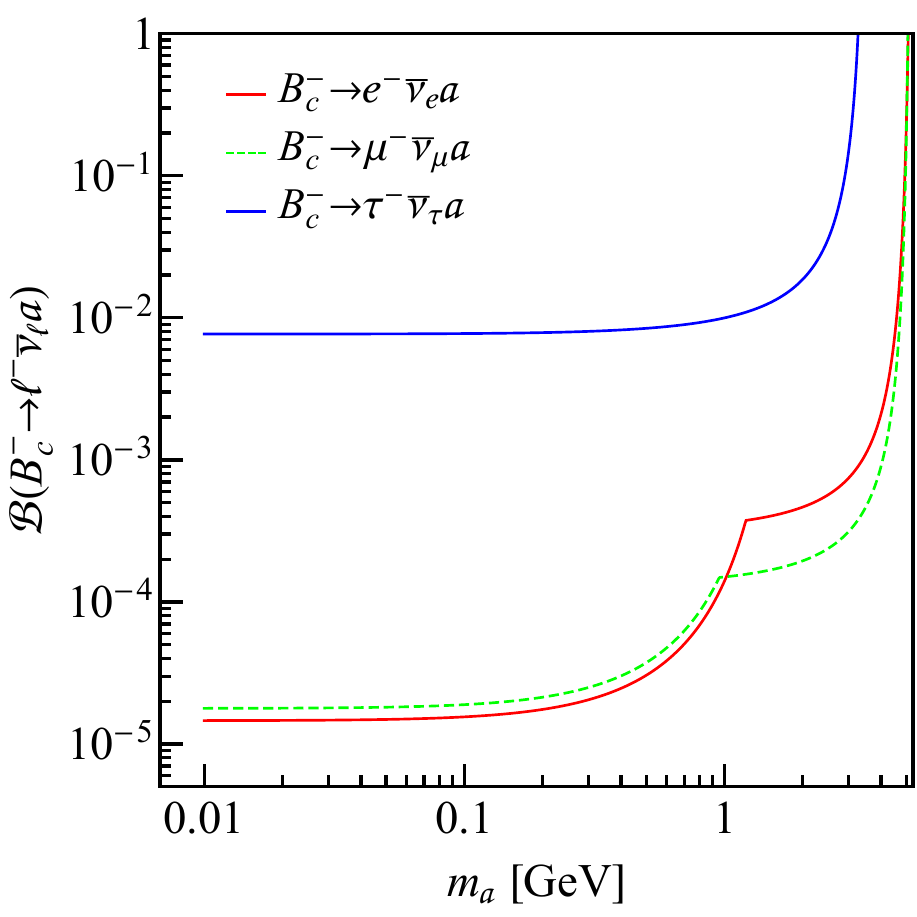}\label{c}
	}
	\quad
	\caption{(color online) The upper limits of $\mathcal{B}(B_c^- \to \ell^- \bar{\nu}_\ell a)$ in scenario 1.}\label{ts}
\end{figure}

\subsection{Scenario 2}

 In this scenario, we consider that the ALP couples with quarks but not with leptons. Two assumptions are made. First, all the coupling constants are assumed to be real numbers, and no additional relative phase is introduced. Second, the coupling constants of the ALP and light quarks, denoted by $c_q$, are assumed to be equal, while for those of heavy quarks, $c_Q$ ($Q=c,~b$) may have different values. We let $c_q$ be a free parameter, then we use the experimental results of $B$ and $D_{(s)}$ mesons to provide constraints on $c_b$ and $c_c$. Specifically, we first write the following expression:
\begin{equation}\label{bc1}
\begin{aligned}
\delta\Gamma&\equiv\Gamma(h^- \to\ell^-\slashed E) - \Gamma(h^- \to
\ell^- \bar{\nu}_\ell) \\
&\ge\Gamma(h^- \to \ell^- \bar{\nu}_\ell a)=\frac{c^2_Q}{f_a^2} \tilde\Gamma_{Q}+\frac{c^2_q}{f_a^2} \tilde\Gamma_{q}+\frac{c_Q c_q}{f_a^2}\tilde\Gamma_{Qq},
\end{aligned}
\end{equation}
where the three terms on the right side of the last equation represent the contributions of Fig.~2(a), 2(b), and their interference, respectively, and $\tilde\Gamma$ denotes the decay width devided by the coupling constant sqaured, which can be calculated theoretically.

\begin{figure}[htbp]
	\centering
	\graphicspath{{figures/}}
	\subfigure[~$m_a=0$~GeV]{
		\includegraphics[width=2.5in,height=2.5in]{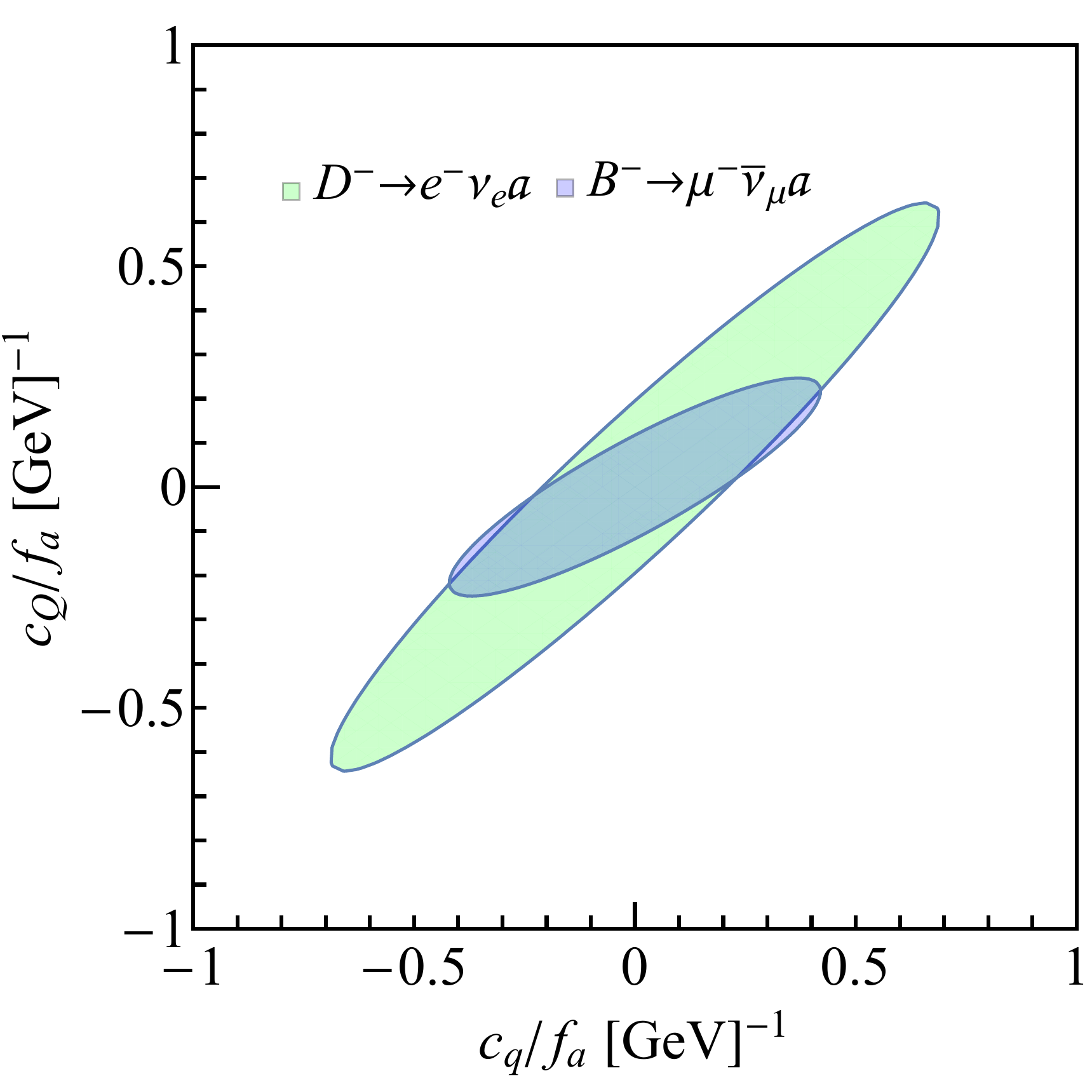}
	}
	\quad
	\subfigure[~$m_a=1.6$~GeV]{
		\includegraphics[width=2.55in,height=2.54in]{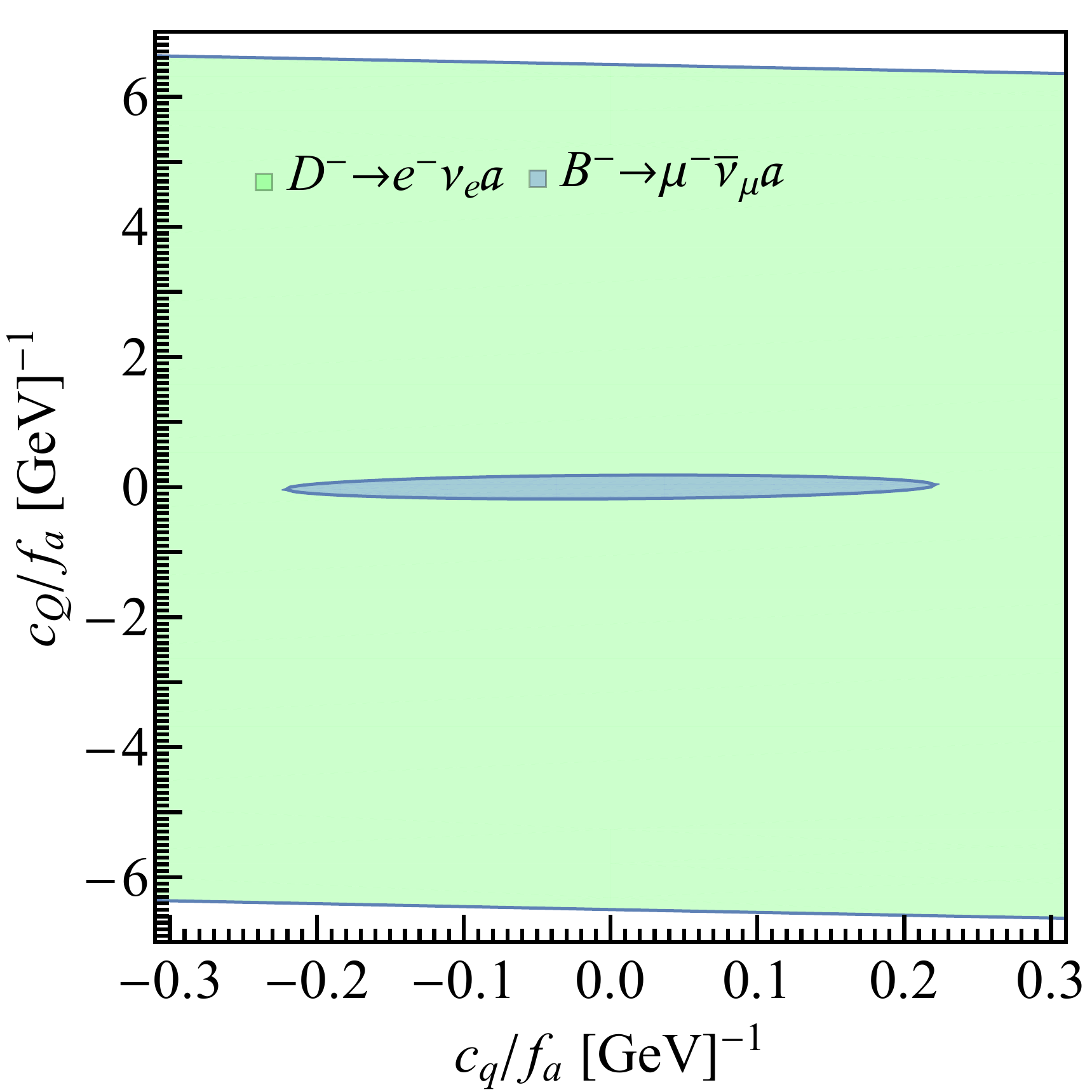}
	}
	\caption{(color online) The experimentally allowed region (inside the ellipse) of the coupling constants. The green and blue areas correspond to $Q=c$ and $Q=b$, respectively.}
	\label{f3}
\end{figure}

The experimentally allowed region for $c_Q$ and $c_q$ is determined by Eq.~(13). We investigate all the decay channels, and find that the decays of $B^-$ and $D^-$ mesons give the most stringent constraint. The area of the parameter space depends on $m_a$. In Fig.~\ref{f3}, we show the results when $m_a$=0, and 1.6~GeV, respectively.  
One can see that when $m=0$ GeV, the two areas are comparable; however, when $m=1.6$ GeV, the area for $Q=c$ (just partially plotted) is larger than that for $Q=b$. 

Next, we scan the experimentally allowed parameter area in Fig.~5 and calculate the limits of $\mathcal{B}(B_c^- \to \ell^- \bar{\nu}_\ell a)$. Because both $c_Q$ and $c_q$ changing the sign does not affect the results, we only need to consider the region with $c_q\ge 0$. Our strategy is as follows: First, by solving Eq.~(13), we obtain the boundary values of $c_Q$ for a fixed $c_q$,
\begin{equation}\label{d3}
\begin{split}
C_{Q(max)} &= \frac{-C_q \tilde\Gamma_{Qq} + \sqrt{(C_q \tilde\Gamma_{Qq})^2-4 \tilde\Gamma_{Q} (C_q^2 \tilde\Gamma_{q} -\delta\Gamma)}}{2 \tilde\Gamma_{Q}}, \\
C_{Q(min)} &= \frac{-C_q \tilde\Gamma_{Qq} - \sqrt{(C_q \tilde\Gamma_{Qq})^2-4 \tilde\Gamma_{Q} (C_q^2 \tilde\Gamma_{q} -\delta\Gamma)}}{2 \tilde\Gamma_{Q}}, 
\end{split}
\end{equation}
 where we have defined $C_{Q(q)}=c_{Q(q)}/f_a$. The condition 
$ C_q^2 \le \frac{4 \tilde\Gamma_Q \delta\Gamma}{4 \tilde\Gamma_Q \tilde\Gamma_{q}-\tilde\Gamma_{Qq}^2}$ should be satisfied to make sure the quantity under the square root nonnegative. Second, we scan $C_b$ and $C_c$ together for a fixed $C_q$ whose allowed region is determined by the small ellipse in Fig.~5.
 
Finally, we calculate the upper limits of $\mathcal{B}(B_c^- \to \ell^- \bar{\nu}_\ell a)$ as functions of $C_q$ with a specific $m_a$. The results are shown in Fig.~6(a) and (c). Here, we only present the result for the positive $C_q$, and that of the negative $C_q$ is symmetrical with it about the vertical axis. As $m_a$ increases, the experimental allowed region of $C_q$ shrinks, and the upper limit of $\mathcal{B}(B_c^- \to \ell^- \bar{\nu}_\ell a)$ roughly increases with $m_a$. This can be understood from Fig.~(5), where as $m_a$ changes from 0 GeV to 1.6 GeV, the allowed range of $c_c/f_a$ becomes quite large, resulting in a larger upper limit. The interesting thing is that for some range of $c_q$, a nonvanishing lower limit of the branching fraction exists(see Fig.~6(b) and (d)). From Fig.~5, we can see that the two ellipses are tilted; therefore, at some $c_q$, $c_b$ and $c_c$ cannot be zero simultaneously, which leads to the nonvanishing $\mathcal{B}(B_c^- \to \ell^- \bar{\nu}_\ell a)$.
\begin{figure}[htbp]
	\centering
	\graphicspath{{figures/}}
	\subfigure[~upper limit with $\ell=e$]{
		\includegraphics[width=2.85in,height=2.85in]{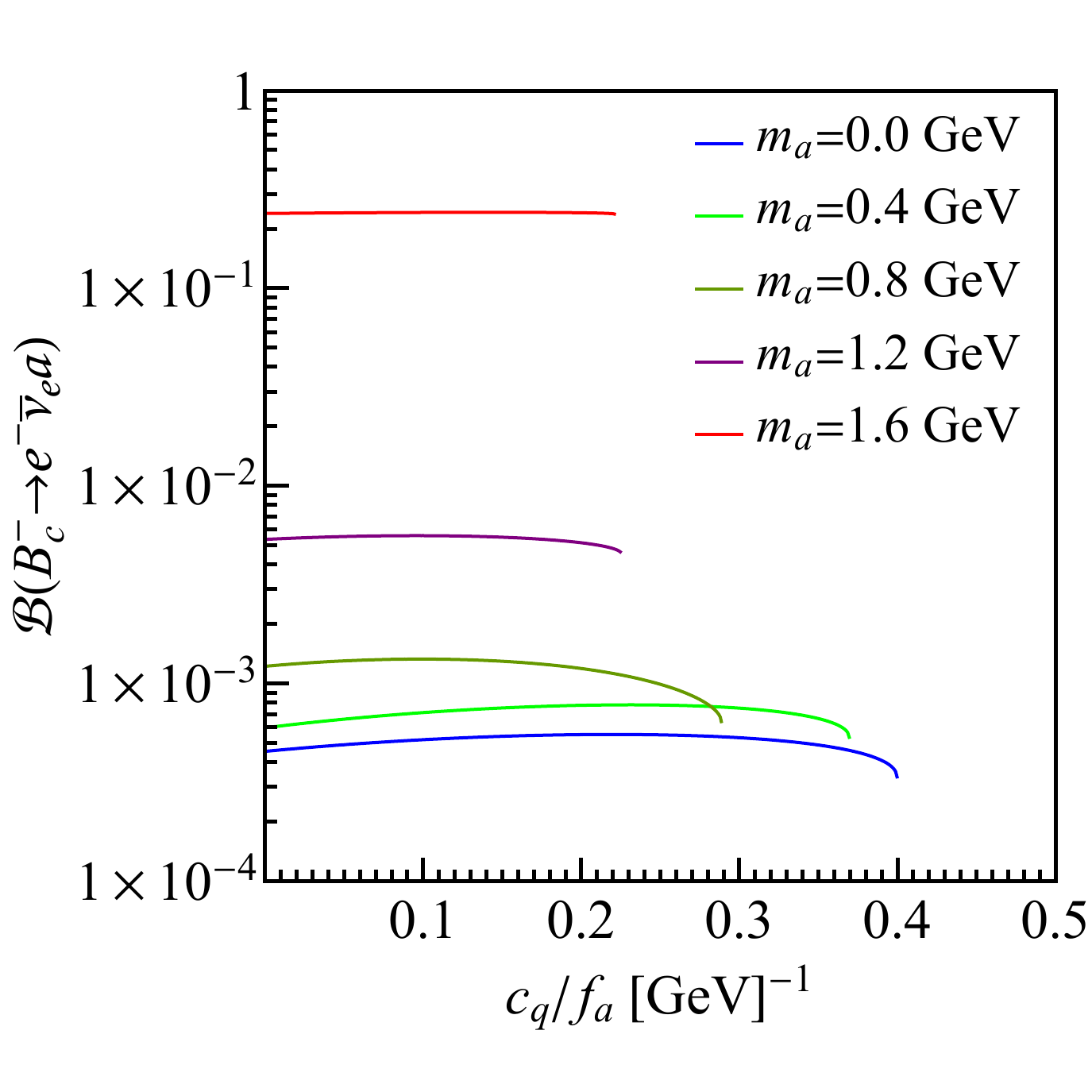}\label{t5a}
	}
	\quad
	\subfigure[~lower limit with $\ell=e$]{
		\includegraphics[width=2.85in,height=2.85in]{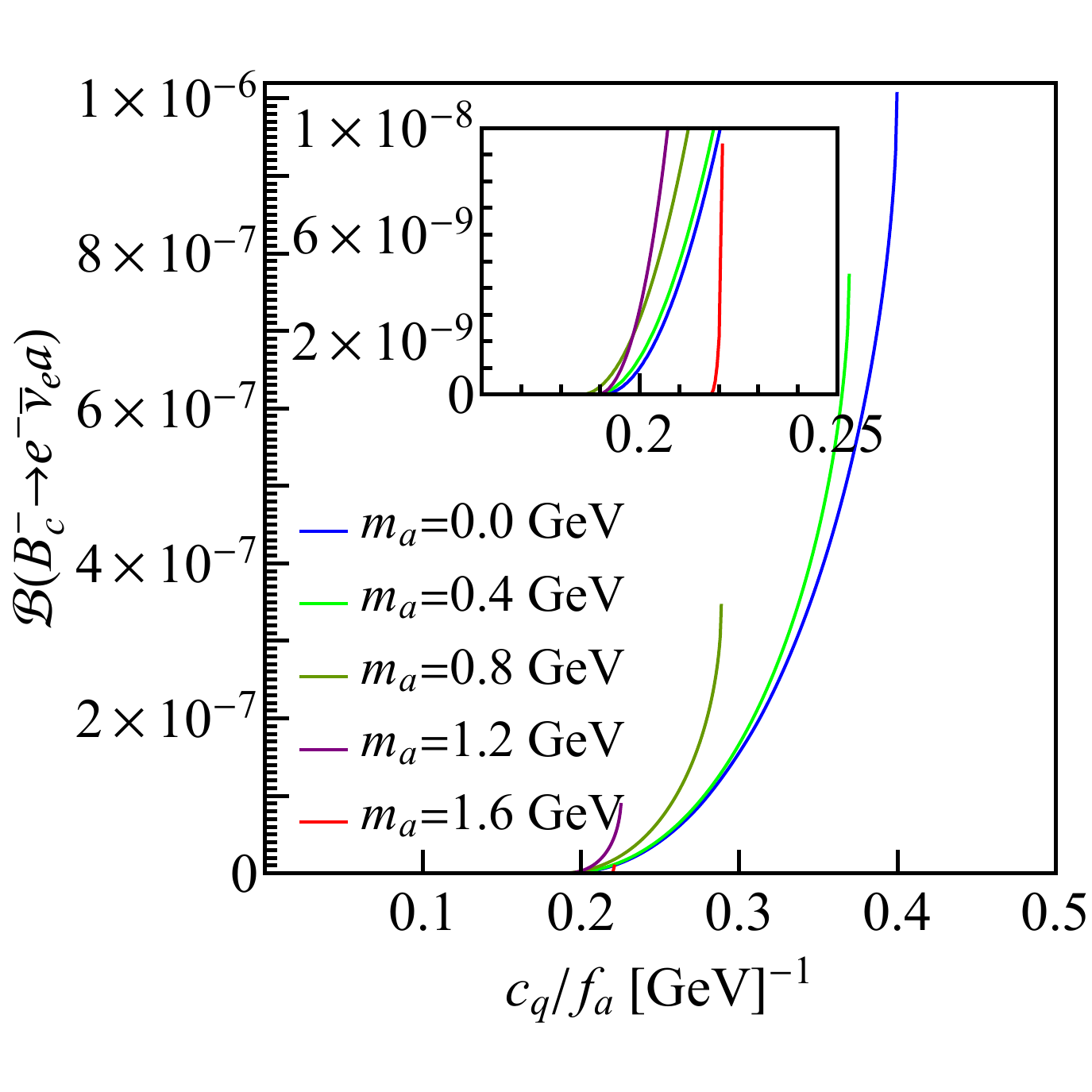}\label{t5b}
	}
	\quad
	\subfigure[~upper limit with $\ell=\tau$]{
		\includegraphics[width=2.85in,height=2.85in]{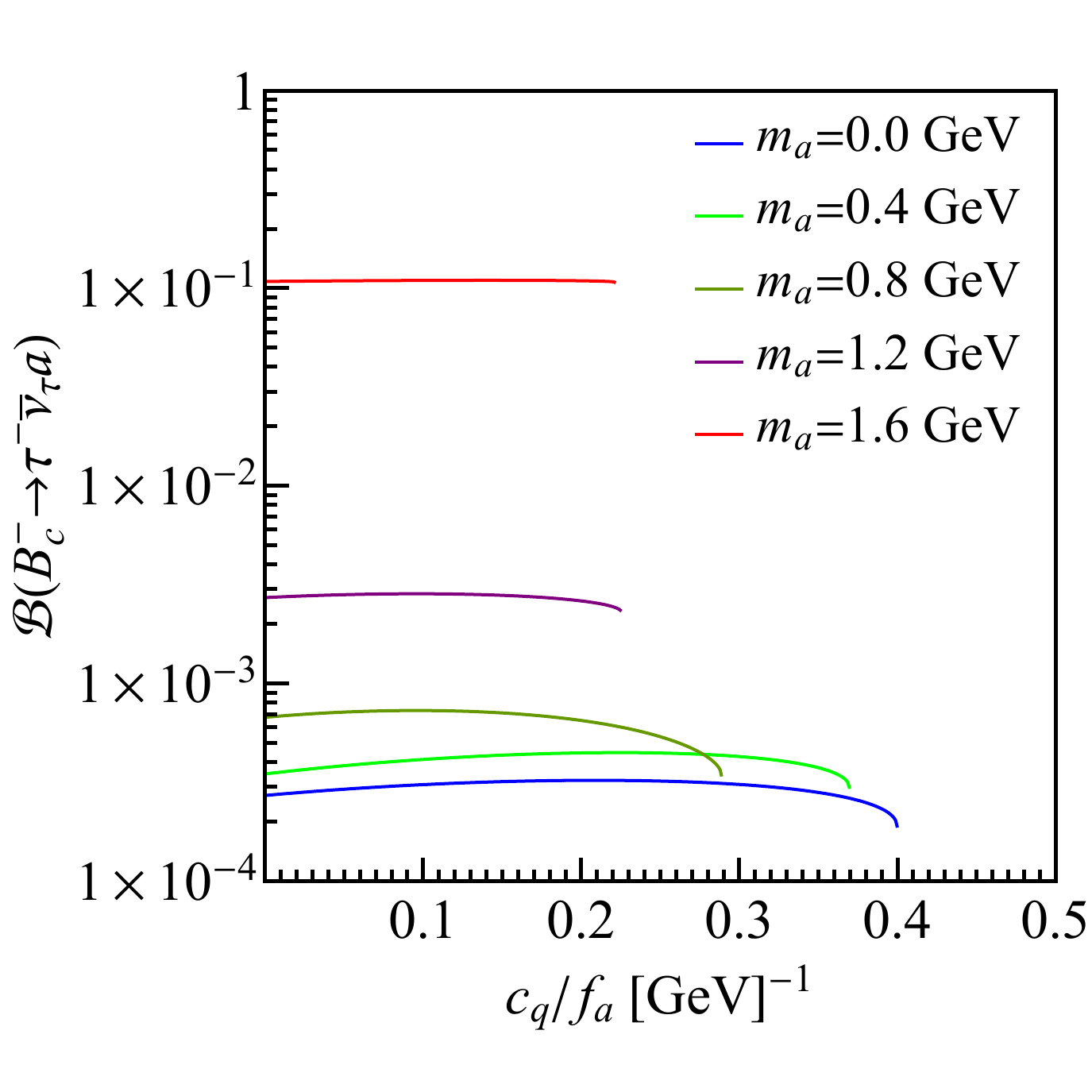}\label{t5e}
	}
	\quad
	\subfigure[~lower limit with $\ell=\tau$]{
		\includegraphics[width=2.85in,height=2.85in]{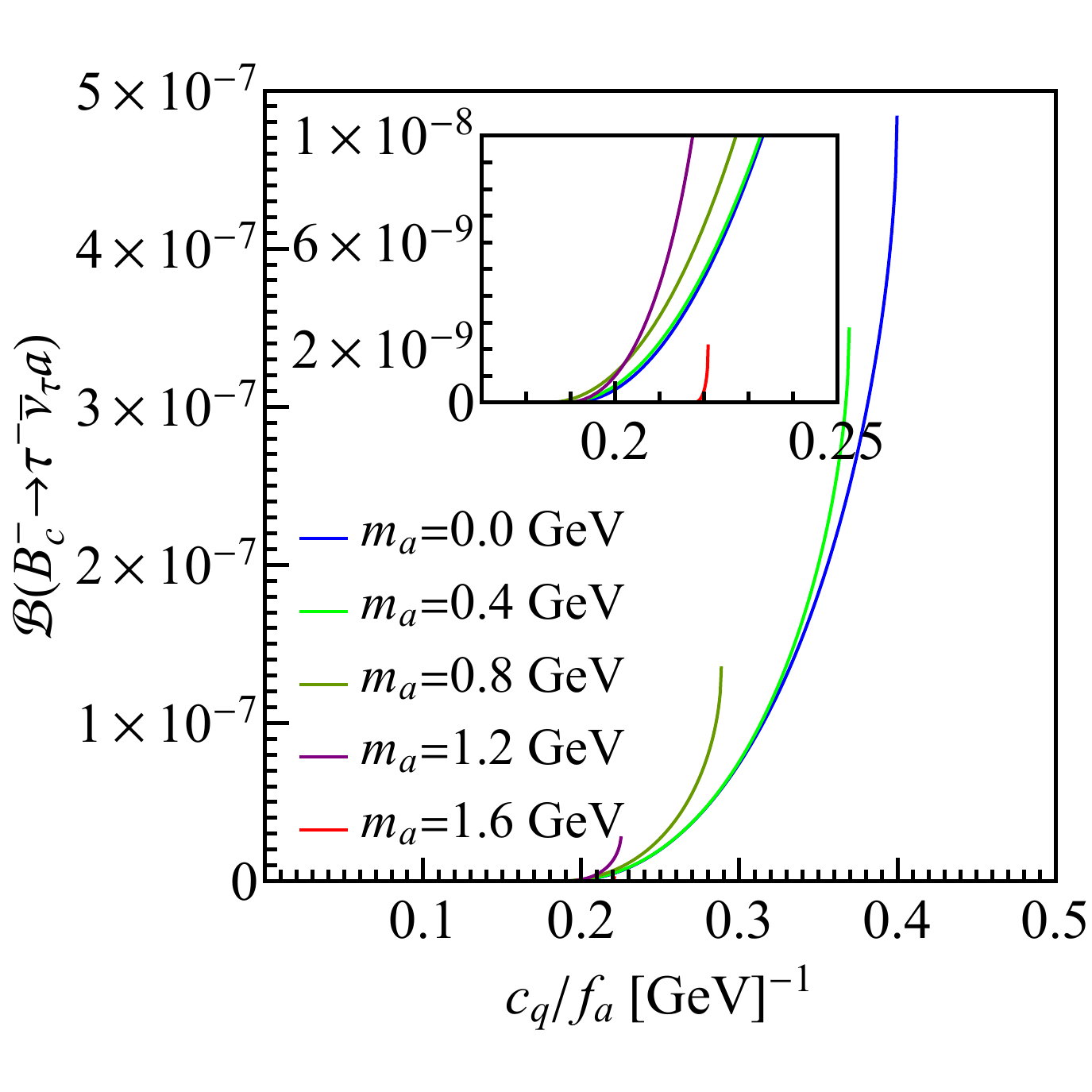}\label{t5f}
	}
	\caption{(color online) The upper and lower limits of $\mathcal B(B_c \rightarrow \ell \bar{\nu}_\ell a)$ in scenario 2.}\label{t5}
\end{figure}

\subsection{Scenario 3}

In this scenario, we assume that the ALP couples with all the charged fermions with the same coupling constant $c_q=c_Q=c_l=c$, the upper limits of which, as functions of $m_a$, are shown in Fig.~7(a). We can see the $B^-\to\mu\bar\nu_\mu a$ channel provides the most stringent restriction. We also notice that the curves have undulation around 1 GeV. We take the $B^-\to\mu\bar\nu_\mu a$ channel as an example to understand why this happened, and plot the contributions of three Feynman diagrams in Fig.~2 and the corresponding interference terms. The effect of the lepton can be neglected owing to its small mass, while the coupling between the ALP and quarks provides the main contribution, especially the interference term, which has a sizeable negative value when $m_a<2$ GeV. This causes the total result to be undulant around $m_a=1$ GeV, which is also transferred to the coupling constant. 

Using the upper limit of the coupling constant given by the $B^-\to\mu\bar\nu_\mu a$ channel, we obtain the constraints of the branching fraction of $B_c^- \to \ell^- \bar{\nu}_{\ell} a$, which are shown in Fig.~8(a). We can see, for $\mathcal{B}(B_c^- \to e^- \bar{\nu}_e a)$ and $\mathcal{B}(B_c^- \to \mu^- \bar{\nu}_{\mu} a)$,that the upper limit is approximately $10^{-6}$ when $m_a$ is less than 1 GeV, and then it keeps increasing until it is larger than one. This is because the phase space of the $B$ decay is small compared with that of the $B_c^-$ decay when $\ell=e$ or $\mu$. For $\mathcal{B}(B_c^- \to \tau^- \bar{\nu}_e a)$, the upper limit is approximately $10^{-5}$ when $m_a<1$ GeV, and then fluctuates with the peak value of $5.4\times 10^{-4}$ at $m_a=3.4$ GeV. Finally, the branching fractions reach zero owing to the vanishing phase space. Comparing these results with the branching fractions of $B_c^-\to\ell^-\bar\nu_\ell$ in the SM is noteworthy. Using Eq.~(1) and (10) we get $\mathcal B(B_c^-\to\ell\bar\nu_\ell)=2.2\times 10^{-9},~9.5\times 10^{-5}$, and $2.2\times 10^{-2}$ for $\ell=e,~\mu$, and $\tau$, respectively. One can see that $\mathcal B(B_c^-\to e^-\bar\nu_e)$ is much smaller than the upper limit of $\mathcal B(B_c^-\to e^-\bar\nu_e a)$ for the chiral suppression; conversely, $\mathcal B(B_c^-\to \tau^-\bar\nu_\tau)$ is much larger than the upper limit of $\mathcal B(B_c^-\to \tau^-\bar\nu_\tau a)$. 
\begin{figure}[htbp]
	\centering
	\graphicspath{{figures/}}
	\subfigure[]{
		\includegraphics[width=2.45in,height=2.45in]{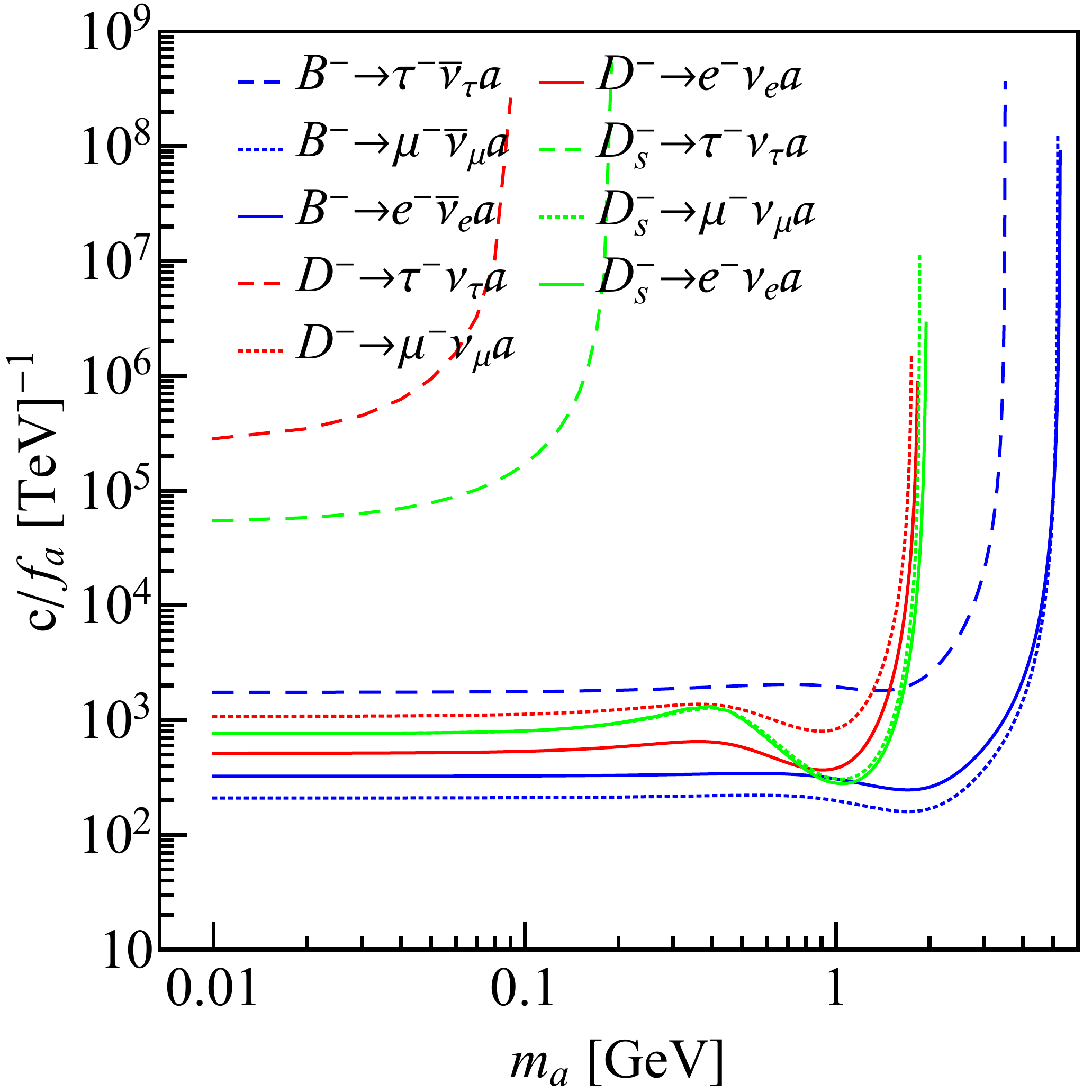}\label{t6a}
	}
	\hspace{1cm}
	\subfigure[]{
		\includegraphics[width=2.45in,height=2.45in]{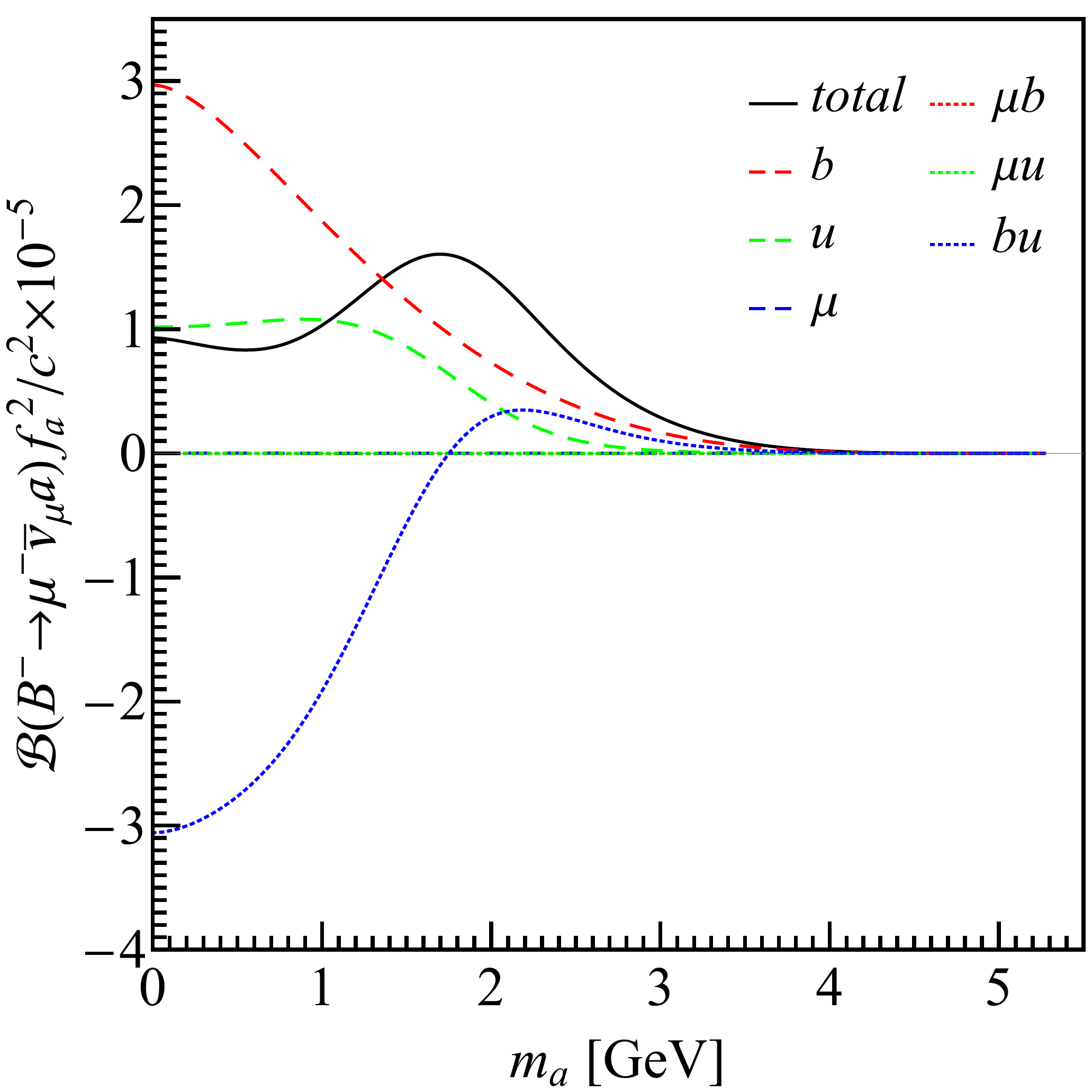}\label{t6b}
	}
	\caption{(color online) (a) shows the upper limits of the coupling constants in scenario 3. (b) Shows the contributions to $\mathcal{B}(B^- \to e^- \bar{\nu}_e a) f_a^2 /c_{i}^{2}$ by each Feynman diagram and the  interference terms.}\label{t6}
\end{figure}

In Fig.~8(b),~(c), and (d), we present the upper limits of the differential branching fractions as functions of the lepton energy. For $\ell$ being $e$ or $\mu$, the results are similar owing to the small lepton mass; however, for $\ell=\tau$, the distribution shape is quite different. We can see that the peak values shrink as $m_a$ increases. Experimentally, the detection of the unmonoenergetic charged lepton in the decay $h^-\to\ell^-\slashed E$ may indicate the existence of ALP, as it has the largest experimentally allowed probability to find the charged lepton around the peak.  
\begin{figure}[htbp]
	\centering
	\graphicspath{{figures/}}
	\subfigure[]{
		\includegraphics[width=2.5in,height=2.5in]{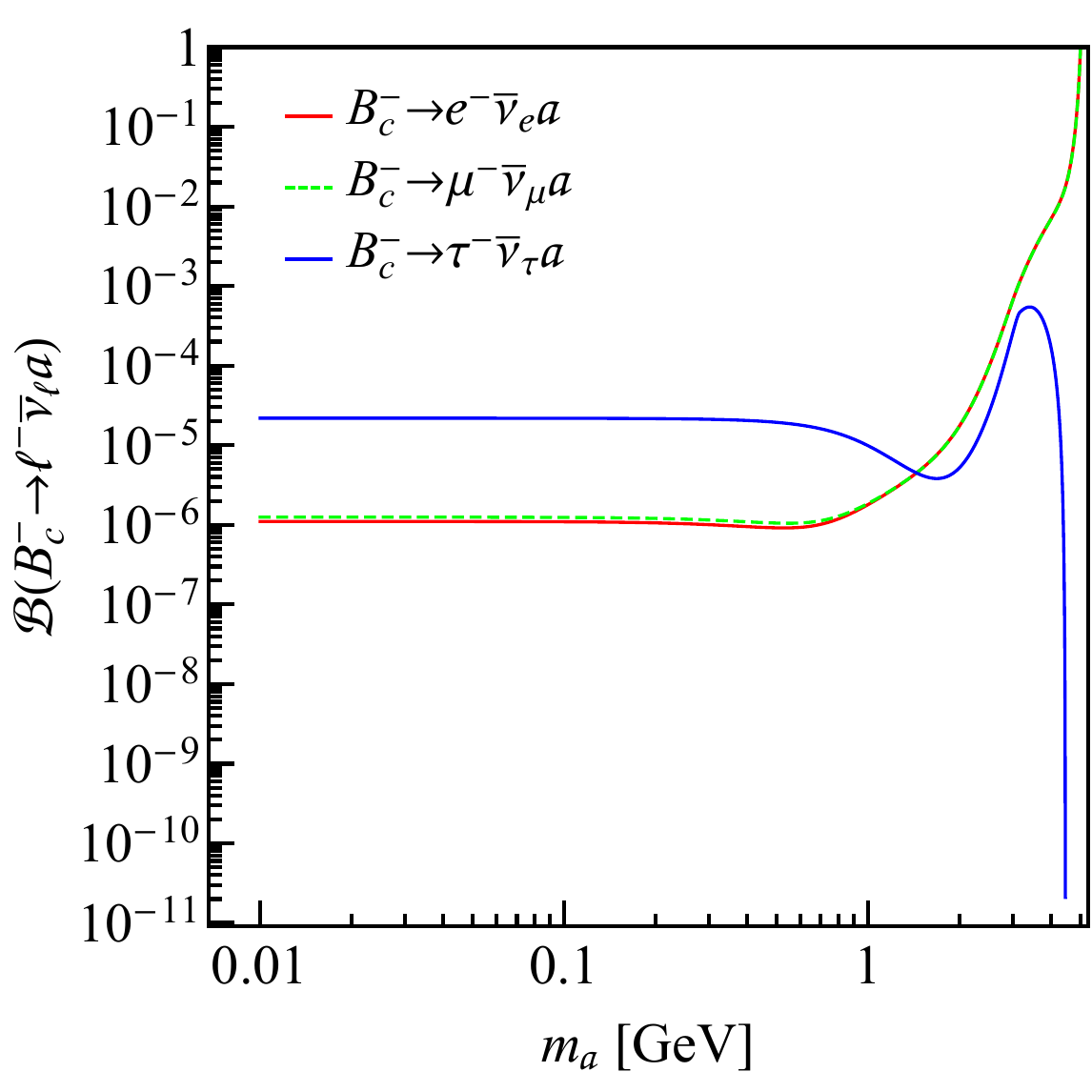}\label{t7a}
	}
	\quad
	\subfigure[]{
		\includegraphics[width=2.65in,height=2.63in]{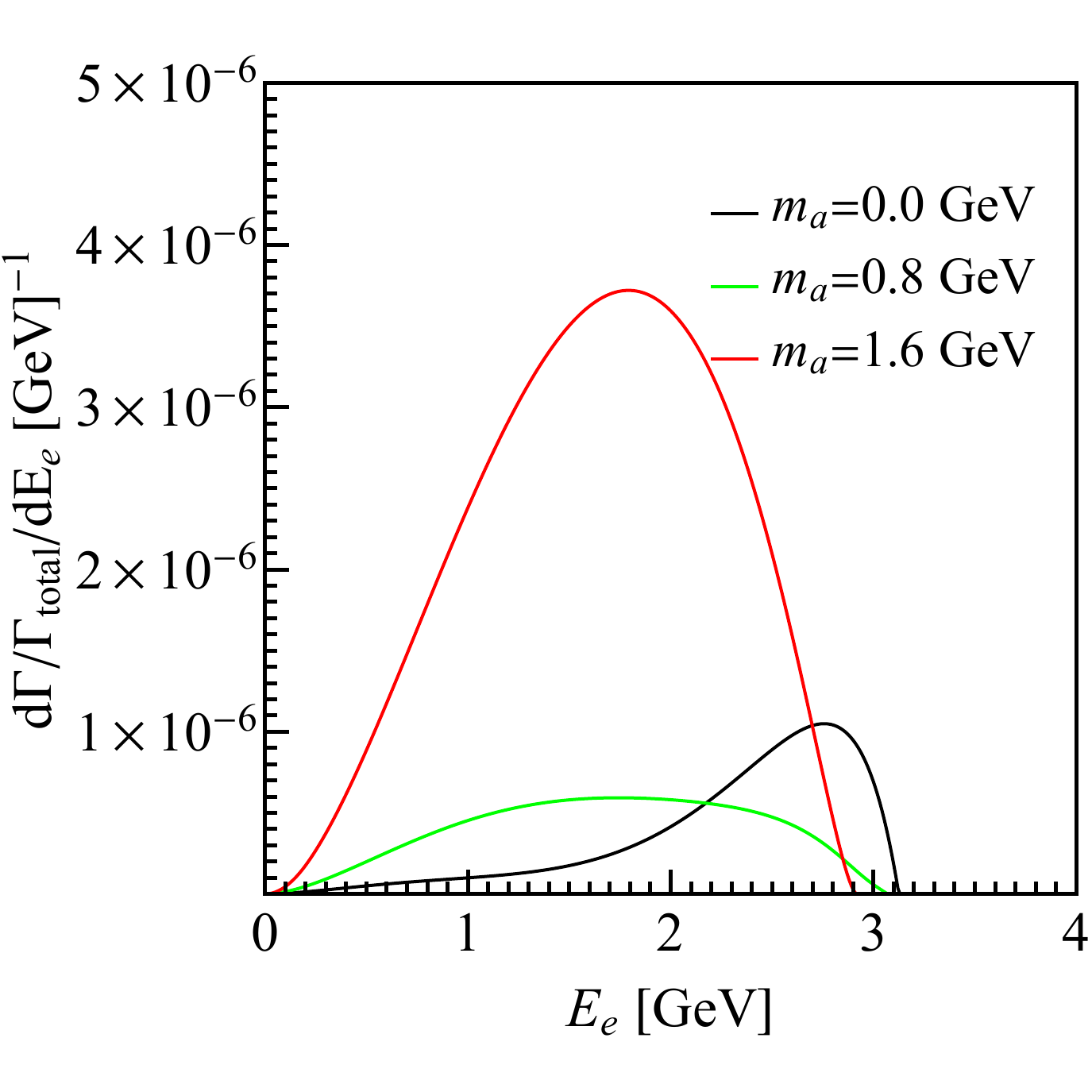}\label{t7b}
	}
	\subfigure[]{
		\includegraphics[width=2.65in,height=2.63in]{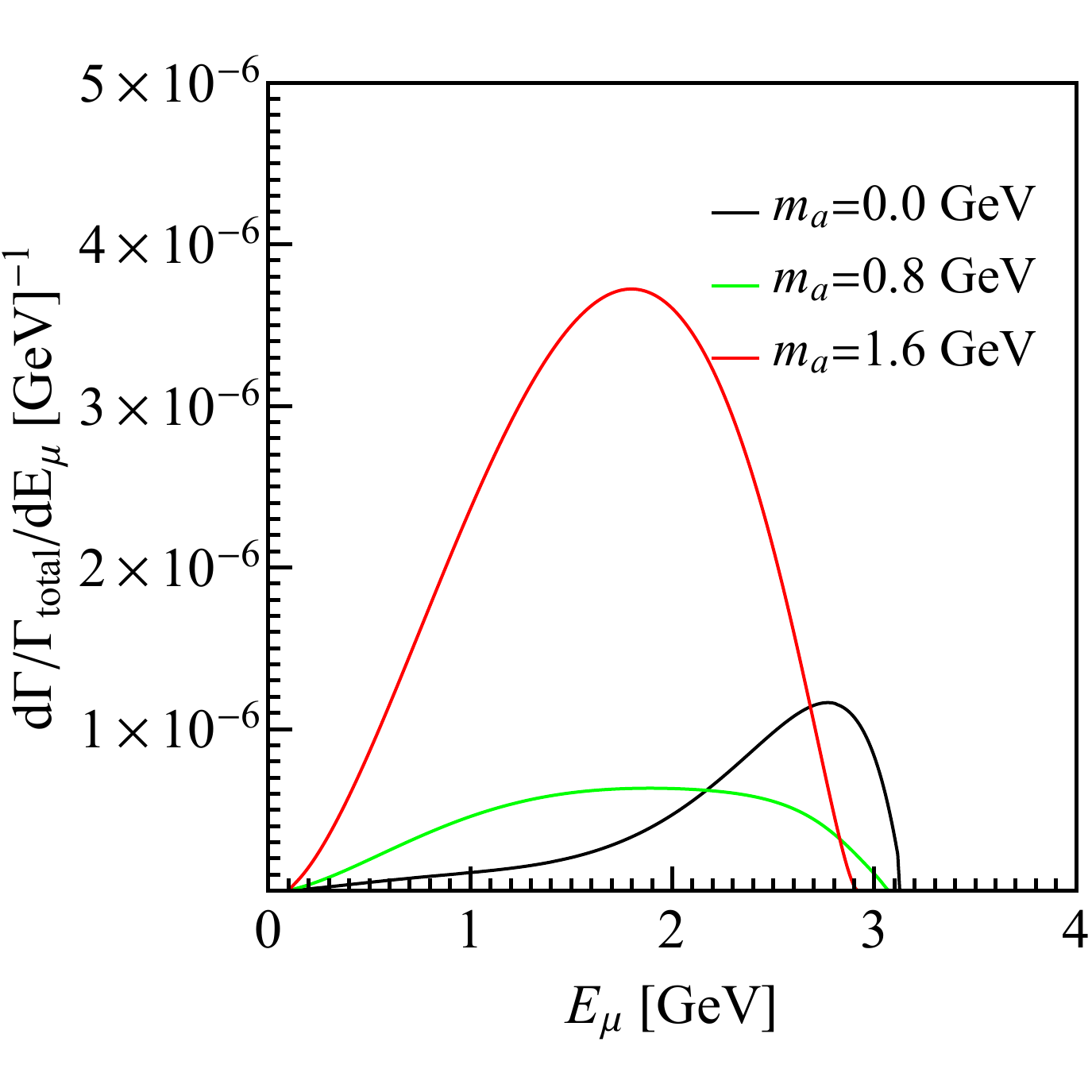}\label{t7c}
	}
	\quad
	\subfigure[]{
		\includegraphics[width=2.65in,height=2.63in]{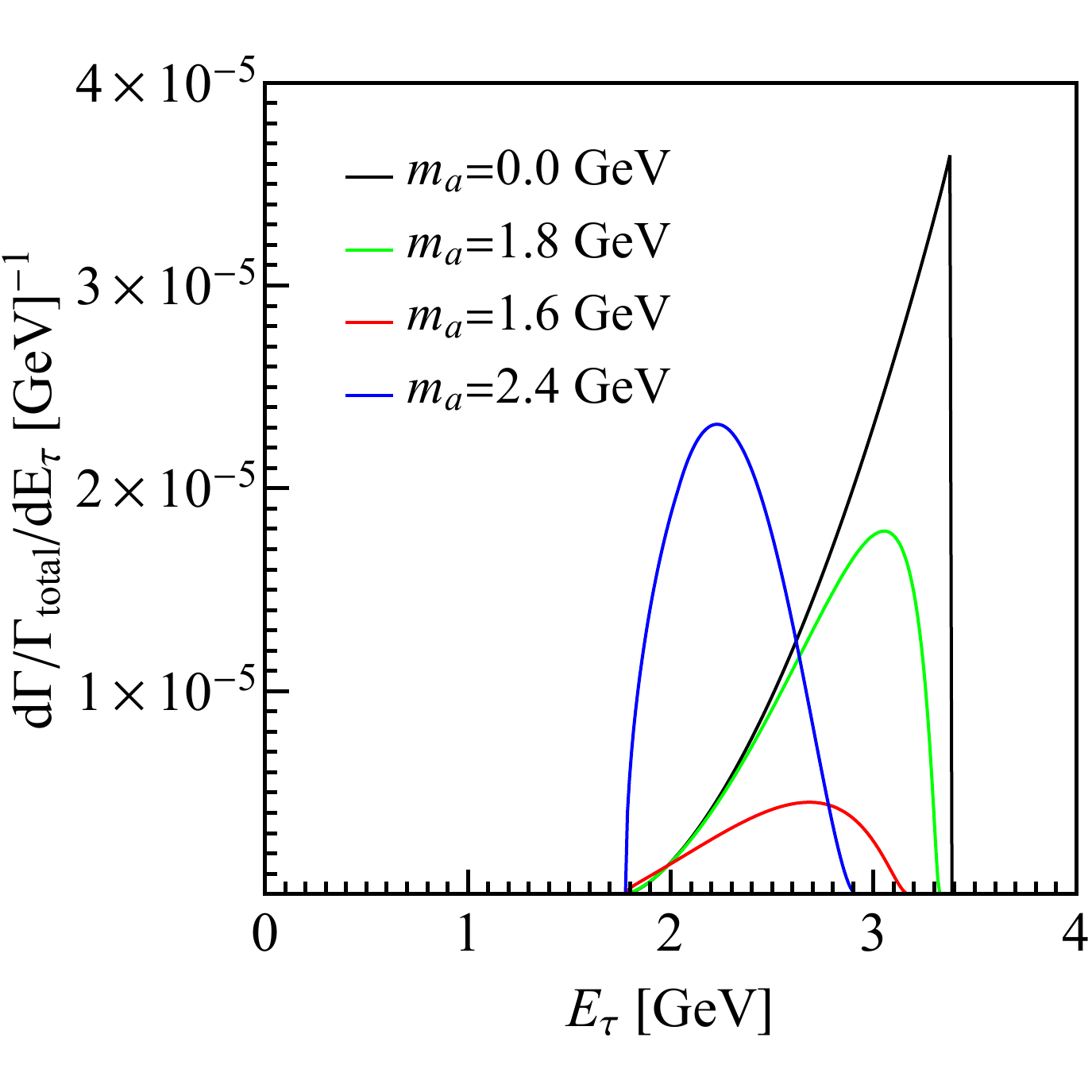}\label{t7d}
	}
	\caption{(color online) (a) shows the upper limit of $ \mathcal{B}(B_c^- \to \ell^- \bar{\nu}_{
		\ell} a)$ in scenario 3. (b), (c), and (d) show the upper limits of the differential branching fraction with $\ell=e$, $\mu$, and $\tau$, respectively.}\label{t7}
\end{figure}


\section{Summary}
\label{Conc}
In conclusion, we have studied ALP production through the processes $h^- \to \ell^- \bar{\nu}_\ell a$.  The instantaneous BS wave functions of the heavy mesons are applied to compute the branching fractions of such decay channels. We adopt three scenarios, that is, the ALP coupling only to one charged fermion, the ALP coupling only to quarks, and the ALP coupling to all the charged fermions with the same coupling constant. In each scenario, by comparing the theoretical and experimental results, we obtain the upper limits of the coupling constants, which are then used to calculate the upper limits of the branching fractions of the $B_c^- \to \ell^- \bar{\nu}_{\ell} a$ channel. For the second scenario, we also obtain the nonzero lower limit of the branching fraction at some range of $c_q$. This study is expected to be helpful for the future detection of ALPs via heavy meson decays.

\acknowledgments

This work was supported by the National Natural Science Foundation of China (NSFC) under Grants No. 12375085, and No. 12075073. T. Wang was also supported by the Fundamental Research Funds for the Central Universities (project number: 2023FRFK06009).

\bibliographystyle{apsrev4-1}
\bibliography{reference}

\end{document}